\documentclass[aps,prab,reprint,groupedaddress,amsmath,amssymb,floatfix, nofootinbib]{revtex4-2}

\usepackage{graphicx}
\usepackage{dcolumn}
\usepackage{bm}
\usepackage{placeins}

\usepackage{comment}
\usepackage{subfigure}
\usepackage{geometry} 
\usepackage[pagebackref]{hyperref}
\renewcommand*\backref[1]{\ifx#1\relax \else ($\uparrow$ #1) \fi}
\usepackage{booktabs}
\usepackage{multirow}

\usepackage{algorithm}
\usepackage[noend]{algpseudocode}

\usepackage{physics}
\usepackage{siunitx}

\usepackage{array}
\usepackage{tabularx}

\usepackage{amsmath,amssymb,bm}
\usepackage{graphicx}
\usepackage{tikz}
\usetikzlibrary{positioning,arrows.meta,calc,fit,backgrounds}
 
\newcommand{\dgtext}{\huge}    
\newcommand{\dgnote}{\huge}    
\newcommand{\dghead}{\Huge}    
 
\newcommand{\panelpath}{.}   
 
\definecolor{cIn}{HTML}{E3EBF4}
\definecolor{cProc}{HTML}{DDE5EC}
\definecolor{cData}{HTML}{F1ECE1}
\definecolor{cModel}{HTML}{DBE7DE}
\definecolor{cOut}{HTML}{EDEDED}
\definecolor{cEdge}{HTML}{333333}
\definecolor{cGrey}{HTML}{8C8C8C}
 
\tikzset{
  >={Stealth[length=8pt,width=6pt]},
  blk/.style   ={draw=cEdge,line width=0.8pt,rounded corners=3.2pt,
                 align=center,inner sep=6pt,font=\dghead},
  wide/.style  ={blk,text width=112mm,minimum height=34mm},
  inb/.style   ={wide,fill=cIn},
  proc/.style  ={wide,fill=cProc},
  data/.style  ={wide,fill=cData},
  model/.style ={wide,fill=cModel},
  use/.style   ={blk,fill=cOut,text width=56mm,minimum height=40mm,font=\dgtext},
  sm/.style    ={blk,font=\dgtext,minimum height=24mm},
  opt/.style   ={sm,dash pattern=on 4.4pt off 3.6pt},   
  ar/.style    ={->,line width=0.8pt,draw=cEdge},
  pl/.style    ={-,line width=0.8pt,draw=cEdge},
  dar/.style   ={ar,dash pattern=on 4.4pt off 3.6pt},
  dpl/.style   ={pl,dash pattern=on 4.4pt off 3.6pt},
  fb/.style    ={ar,dash pattern=on 4pt off 3pt},
  lead/.style  ={-,line width=0.6pt,draw=cGrey,dash pattern=on 2pt off 2pt},
  lbl/.style   ={font=\dgnote\itshape,text=black!70,align=center},
  pan/.style   ={inner sep=0pt,outer sep=0pt},
}

\begin{document}

\preprint{APS/123-QED}

\title{Flow-based surrogate models for particle tracking} 
\author{Matthias Remta}
\affiliation{CERN, Geneva 1211, Switzerland}
\affiliation{University of Vienna, Vienna 1090, Austria}
\author{Yann Dutheil}
\affiliation{CERN, Geneva 1211, Switzerland}
\author{Francesco Velotti}
\affiliation{CERN, Geneva 1211, Switzerland}

\date{\today}

\begin{abstract}
Particle tracking is a fundamental tool for particle-accelerator design and optimisation. Conventional tracking routines provide high accuracy but are computationally demanding, especially when simulating large particle ensembles or long time spans. As a result, optimising moderate- to high-dimensional parameter spaces is challenging, and real-time surrogate models remain out of reach for many applications.

This contribution introduces a surrogate-modelling approach based on conditional flow matching (CFM). A CFM model is trained on tracking simulations of CERN's Proton Synchrotron (PS) over a 10-dimensional parameter space. The trained model reproduces final phase-space distributions with a median squared maximum mean discrepancy (\(\text{MMD}^2\)) of \num{3e-4} and mean inference time of \SI{0.04}{\second}, a speed-up of three orders of magnitude over conventional tracking.

To capture distribution-dependent dynamics that vanilla CFM cannot represent, we extend the model with cross-attention over the initial particle ensemble (Cross-Attention-CFM) and demonstrate that this extension recovers performance on a space-charge benchmark in the PS, where vanilla CFM degrades.

Finally, we introduce Hybrid-CFM, in which a small number of conventionally-tracked particles are used to inform the model. On the same 10-dimensional PS task, Hybrid-CFM with 100 auxiliary particles trained on 200 distributions matches the vanilla CFM trained on 1500, and improves the worst-case (90th-percentile) \(\text{MMD}^2\) by roughly a factor of four, substantially reducing the upfront cost of building a surrogate.
\end{abstract}

\maketitle

\section{Introduction}\label{sec:intro}

A central objective in accelerator physics is the control of the phase-space distribution of the particle beam. Containing the particles within the aperture of the machine is essential for safe and efficient operation, while more specific requirements on the distribution arise in many situations, such as efficient beam transfer, loss minimisation, focusing at collider interaction points, and the protection of fixed targets~\cite{remta:ipac25-tupb018}. Consequently, simulating the time-evolution of particle distributions is essential for the design, operation and optimisation of modern particle accelerators. In practice, such simulations represent the distribution as a large ensemble of macro-particles and evolve them with particle-tracking routines based on symplectic integration~\cite{Grote:2003ct, Iadarola:2023fuk, kaiser2024cheetah}. These routines provide high accuracy but are computationally demanding, especially when simulating long time spans, large ensembles, or many parameter configurations. As a result, design studies, optimisation campaigns and real-time control loops are bottlenecked by tracking cost.

Three approaches have been developed to alleviate this cost. \emph{Differentiable tracking codes}~\cite{kaiser2024cheetah, cbds-lst1} enable gradient-based optimisation through auto-differentiation. This approach is highly effective for linear or short-section problems but becomes intractable for long-term tracking in circular accelerators, where the computation graph grows prohibitively with the number of turns. \emph{Black-box neural surrogates}, including DeepONet and Fourier neural-operator architectures, have been used to predict beam-envelope statistics and downstream quantities in linear accelerators~\cite{particles8010021}. These models offer large inference speed-ups but struggle with non-linear dynamics for long integration times~\cite{remta2025}. \emph{Structure-preserving surrogates} enforce symplecticity in the architecture. Examples are physics-based networks built on the Taylor expansion of the transfer map~\cite{PhysRevAccelBeams.23.074601}, H{\'e}non-map architectures~\cite{Burby_2021, Huang_2024} and SympNet~\cite{JIN2020166, remta2025}. The strong inductive bias improves accuracy on single-particle dynamics, but iterative turn-by-turn inference limits speed-ups.

What none of these threads addresses simultaneously is (i) modelling the full six-dimensional phase-space distribution -- with or without collective effects -- rather than per-particle or low-dimensional summary maps, (ii) end-to-end differentiability with respect to high-dimensional machine settings at long-term-tracking scales and (iii) sub-second inference. We close this gap with a generative surrogate based on \emph{conditional flow matching} (CFM)~\cite{lipman2023, tong2024}. CFM learns a continuous-time vector field that maps an initial phase-space distribution to the final distribution after tracking, conditioned on the machine settings.

The contributions of this work are threefold:
\begin{enumerate}
    \item We introduce CFM as a surrogate for particle tracking conditioned on machine parameters, and demonstrate near-real-time inference on a 10-dimensional task in the PS.
    \item We extend the model with cross-attention over the initial particle ensemble (CA-CFM) to capture distribution-dependent dynamics that vanilla CFM cannot represent, recovering performance on a space-charge problem.
    \item We introduce \emph{Hybrid-CFM}, in which a small number of conventionally-tracked auxiliary particles is cross-attended to inform the model, achieving up to a $7.5\times$ reduction in upfront tracking-simulation cost at matched median accuracy and up to a $4\times$ improvement in worst-case performance.
\end{enumerate}

Section~\ref{sec:method} introduces CFM, the CA-CFM and Hybrid-CFM extensions, the network architecture, training pipeline and evaluation metric. Section~\ref{sec:results} presents three experiments: a 10-dimensional optimisation task in the PS, a space-charge benchmark, and a data-efficiency study. Section~\ref{sec:discussion} discusses outlier behaviour, compute trade-offs and limitations, and Section~\ref{sec:conclusion} concludes.

\section{Methods}\label{sec:method}

\subsection{Conditional flow matching}\label{sec:cfm}

Flow matching (FM)~\cite{lipman2023} learns to transform samples from a simple initial distribution $p_0$ into samples from a target distribution $p_1$ by integrating a time-dependent vector field $v_t : \mathbb{R}^d \rightarrow \mathbb{R}^d$ over $t \in [0,1]$. The field $v_t$ generates a flow $g_t$ via
\begin{equation}\label{eq:flow}
    \frac{d}{dt}g_t(x) = v_t(g_t(x)), \quad g_0(x) = x,
\end{equation}
which pushes $p_0$ forward to $p_t = [g_t]_* p_0$ through the change-of-variables formula. A neural network $u_t(x;\theta)$ is trained to match $v_t$ by minimising
\begin{equation}\label{eq:lfm}
    \mathcal{L}_\text{FM}(\theta) = \underset{t,\, p_t(x)}{\mathbb{E}}\|u_t(x;\theta) - v_t(x)\|^2.
\end{equation}
However, neither $p_t$ nor $v_t$ is available in closed form in general, rendering $\mathcal{L}_\text{FM}$ intractable~\cite{lipman2023}. Conditioning on sample pairs $(x_0, x_1)$ drawn from a joint distribution $q(x_0, x_1)$ yields the tractable conditional FM (CFM) objective~\cite{lipman2023, tong2024}
\begin{equation}\label{eq:lcfm}
\begin{split}
    \mathcal{L}_\text{CFM}(\theta) = \underset{t,\, q(x_0,x_1),\, p_t(x|x_0,x_1)}{\mathbb{E}} \\
    \|u_t(x;\theta) - v_t(x|x_0,x_1)\|^2,
\end{split}
\end{equation}
which shares the same $\theta$-gradients as $\mathcal{L}_\text{FM}$~\cite{lipman2023}. Following Ref.~\cite{tong2024}, we adopt Gaussian conditional paths
\begin{equation}\label{eq:gaussian-path}
    p_t(x|x_0,x_1) = \mathcal{N}\!\left(x \,\big|\, tx_1 + (1-t)x_0,\, \sigma^2 I\right),
\end{equation}
with $\sigma$ being a small fixed hyperparameter (see Appendix~\ref{app:inference}). The corresponding conditional vector field is the constant
\begin{equation}\label{eq:cond-vfield}
    v_t(x|x_0,x_1) = x_1 - x_0.
\end{equation}

We apply CFM to learn the map between initial and final beam distributions in phase space. The endpoints $p_0$ and $p_1$ are represented by ensembles $(x_0^{(i)})_{i=1}^N$ and $(x_1^{(i)})_{i=1}^N$ of 6-dimensional phase-space coordinates before and after tracking, respectively. Because the CFM objective does not require paired samples, the particle correspondence between $x_0^{(i)}$ and $x_1^{(i)}$ is discarded during training. We condition the learned vector field on the machine settings via an additional input $c \in \mathbb{R}^{d_c}$, writing $u_t(x;\theta,c)$, so that a single trained model covers the entire parameter space.

\subsection{Cross-Attention-CFM}\label{sec:ca-cfm}

Vanilla CFM assumes that, for fixed $c$, the dynamics depend on each particle's initial coordinates but not on the rest of the initial ensemble. This assumption fails for collective effects, where every particle's trajectory depends on the global distribution. For initial distributions parameterised by a small set of scalars this dependence could in principle be absorbed into $c$. For non-parametric distributions, however, there is no obvious way to do so, and we instead extend the model with cross-attention~\cite{vaswani2017} between the initial ensemble and the network's latent representation. Each initial coordinate $x_0^{(i)}$ is embedded into the network's hidden dimension and serves as a key/value token. The latent representations inside the backbone act as queries. We refer to this architecture as \emph{Cross-Attention-CFM} (CA-CFM). Because no positional encoding is applied, cross-attention is permutation-invariant in the token dimension and naturally handles ensembles of arbitrary and varying size.

\subsection{Hybrid-CFM}\label{sec:hybrid}

In high-dimensional non-linear problems the required volume of training data can become a limiting factor for surrogate models. We mitigate this curse of dimensionality by combining the CFM model with computationally cheap conventional simulations. Concretely, for each query distribution we run conventional single-particle tracking for a small auxiliary ensemble of $N_\text{aux} \ll N$ particles drawn from the same initial distribution as the main ensemble, and present the resulting \emph{final} phase-space coordinates $\{x_1^{(\text{aux},j)}\}_{j=1}^{N_\text{aux}}$ as cross-attention tokens to the network. We call this architecture \emph{Hybrid-CFM}.

\subsection{Surrogate-modelling pipeline}\label{sec:pipeline}

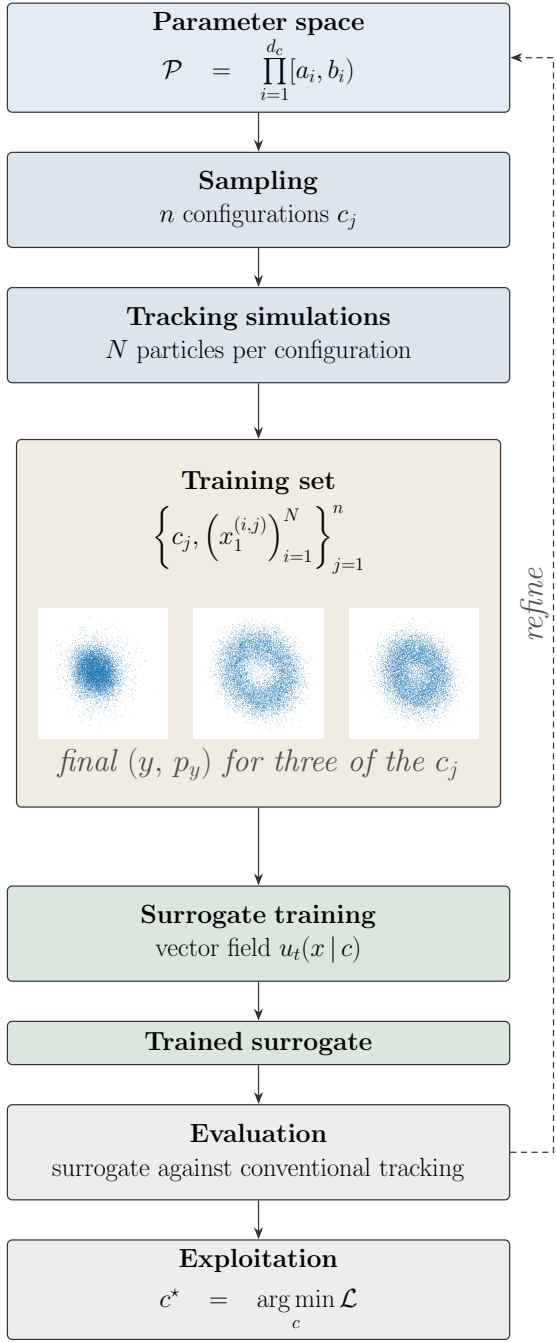
\begin{figure}
\resizebox{\columnwidth}{!}{%
\begin{tikzpicture}[x=2mm,y=2mm, node distance=9mm,
                    blk/.append style={font=\LARGE},
                    wide/.append style={text width=112mm, minimum height=22mm}]
 
  \node[inb]              (P) {\textbf{Parameter space} \\ $\mathcal{P}=\prod\limits_{i=1}^{d_c} [a_i,b_i)$};
  \node[proc, below=of P] (S) {\textbf{Sampling} \\ $n$ configurations $c_j$};
  \node[proc, below=of S] (T) {\textbf{Tracking simulations}\\
                               $N$ particles per configuration};
 
  \node[blk,draw=none,fill=none, below=18mm of T] (D) {\textbf{Training set}\\[6pt]
        $\left\{c_j,\left(x_1^{(i,j)}\right)_{i=1}^{N}\right\}_{j=1}^{n}$};
 
  \node[pan,below=6mm of D]   (pc2) {\includegraphics[width=30mm]{\panelpath/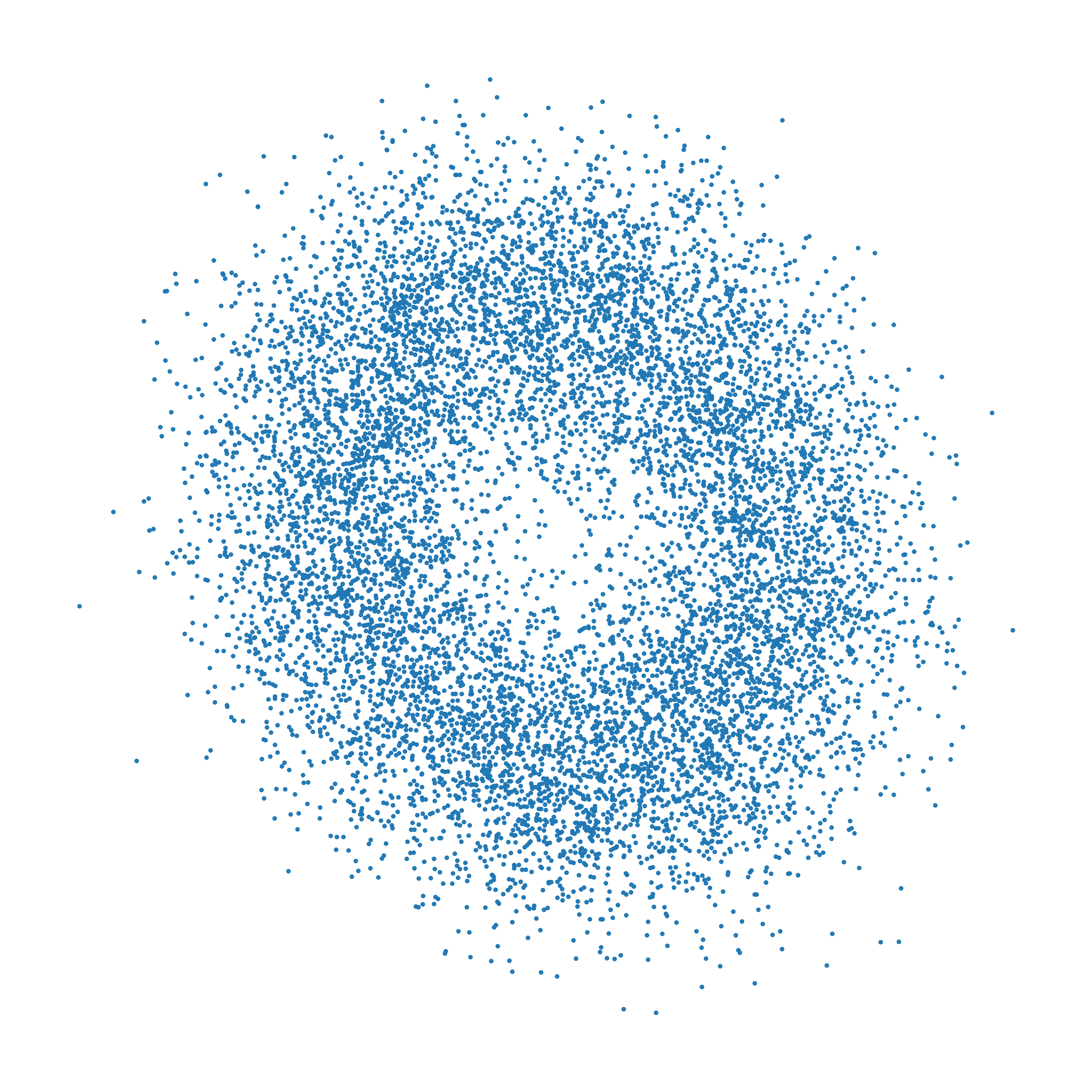}};
  \node[pan,left =6mm of pc2] (pc1) {\includegraphics[width=30mm]{\panelpath/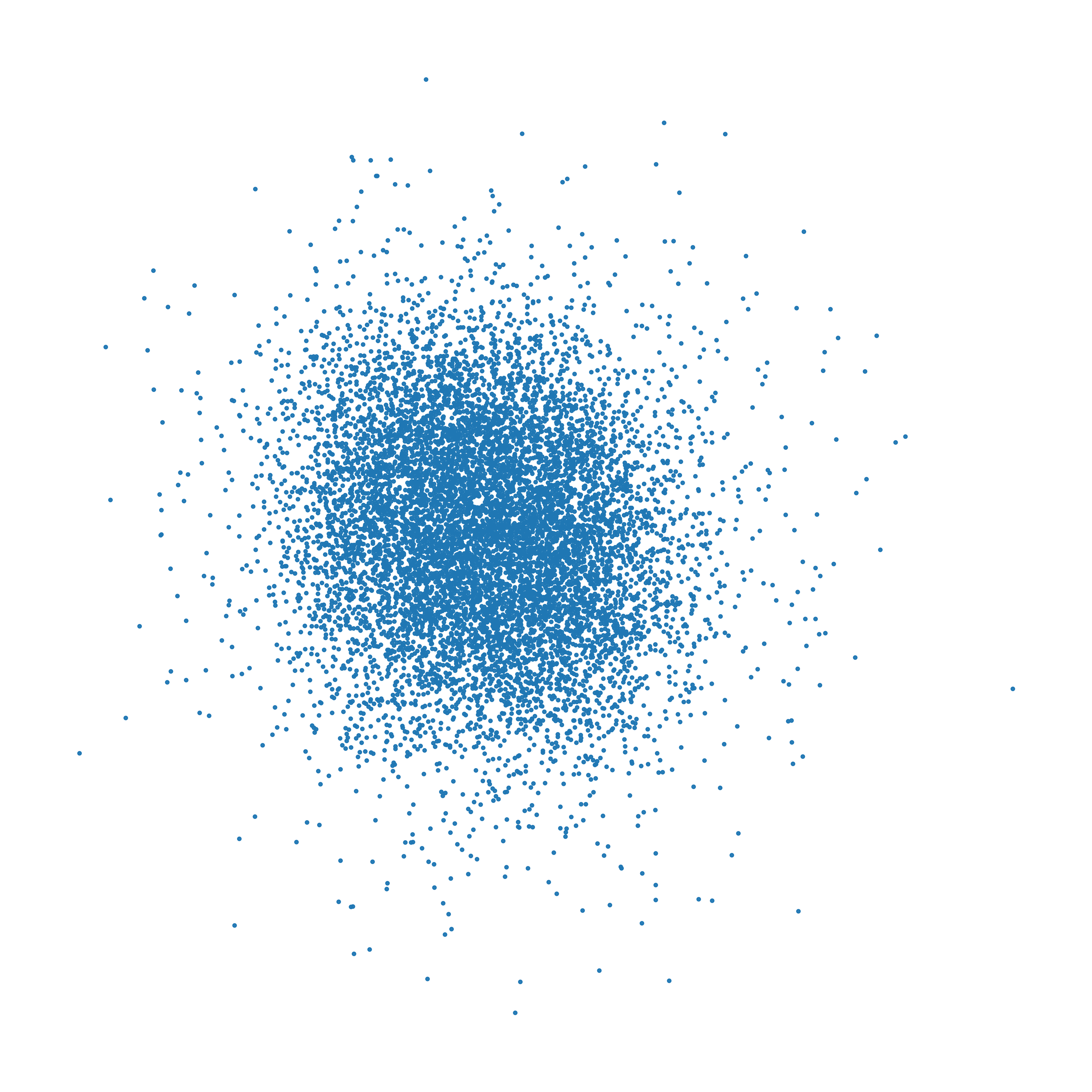}};
  \node[pan,right=6mm of pc2] (pc3) {\includegraphics[width=30mm]{\panelpath/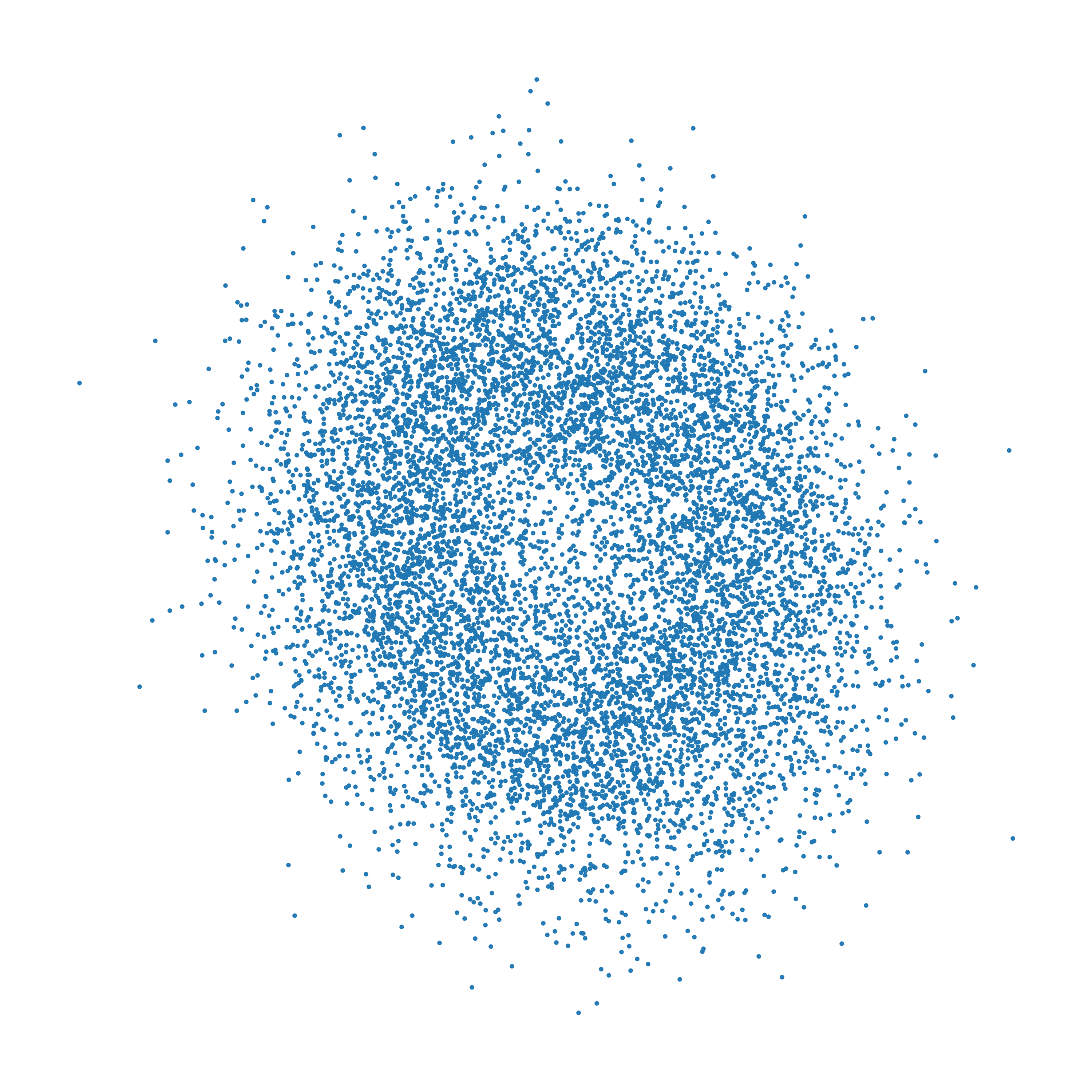}};
  \node[lbl,below=1mm of pc2] (plab) {final $(y,\, p_y)$ for three of the $c_j$};

  \begin{scope}[on background layer]
    \node[draw=cEdge,line width=0.8pt,rounded corners=3.2pt,fill=cData,
          inner sep=5mm,fit=(D)(pc1)(pc3)(plab)] (DG) {};
  \end{scope}
 
  \node[model,below=18mm of DG] (M) {\textbf{Surrogate training}\\
                                     vector field $u_t(x\,|\,c)$};
  \node[model,below=of M,minimum height=10mm] (G) {\textbf{Trained surrogate}};
 
  \foreach \a/\b in {P/S,S/T} \draw[ar] (\a) -- (\b);
  \draw[ar] (T) -- (DG);          
  \draw[ar] (DG) -- (M);          
  \draw[ar] (M) -- (G);
 
  \node[wide,fill=cOut, below=of G] (E)
       {\textbf{Evaluation}\\ surrogate against conventional tracking};
  \node[wide,fill=cOut, below=of E] (X)
       {\textbf{Exploitation}\\[4pt] $c^\star = \operatorname*{arg\,min}\limits_c \mathcal{L}$};
 
  \draw[ar] (G) -- (E);
  \draw[ar] (E) -- (X);
 
  \coordinate (fb1) at ([xshift=10mm]E.east);
  \draw[fb] (E.east) -- (fb1) |- (P.east);
  \node[lbl,rotate=90,anchor=south] at ($(fb1)!0.5!(fb1|-P.east)$) {refine};
 
\end{tikzpicture}}
\caption{Flowchart of the surrogate pipeline.}
\label{fig:pipeline}
\end{figure}

A surrogate model for a given application is built as follows. We first define a parameter space
\begin{equation}\label{eq:pspace}
    \mathcal{P} = \prod_{i=1}^{d_c} [a_i, b_i),
\end{equation}
encoding for example initial-distribution parameters, magnet strengths, and the voltages and frequencies of radio-frequency cavities. We then draw $n$ points from $\mathcal{P}$, using grid sampling for low-dimensional spaces and scrambled Sobol sampling~\cite{OWEN1998466} otherwise, which produces a low-discrepancy sequence in the hypercube $[0,1)^{d_c}$ and covers $\mathcal{P}$ efficiently. For each of the $n$ configurations we perform a conventional tracking simulation, evolving an ensemble of $N$ particles chosen to describe the distribution adequately. The resulting dataset is used to train the CFM model (or one of its extensions). After training, the model can be used in two regimes: (i) as a fast tracking surrogate during forward design studies, and (ii) as a differentiable map from machine settings to the final distribution, enabling gradient-based search for an optimal $c^\star \in \mathcal{P}$ with respect to any differentiable objective function \(\mathcal{L}\) that measures the distance to the target distribution. Figure~\ref{fig:pipeline} shows a flowchart of the pipeline.

\subsection{Network architecture}\label{sec:arch}

The vector field $u_t$ is parametrised by a residual network~\cite{he2016} with ten blocks, each consisting of two fully-connected layers bypassed by a skip connection. The network takes as input the seven-dimensional concatenation of the phase-space coordinate $x \in \mathbb{R}^6$ and the scalar time $t$, and outputs the corresponding six-dimensional vector field. The conditioning vector $c \in \mathbb{R}^{d_c}$ enters as an additional input to each residual block through a separate channel, in addition to being injected once at the network entry point. For CA-CFM and Hybrid-CFM (Sections~\ref{sec:ca-cfm} and~\ref{sec:hybrid}), cross-attention modules are inserted into residual blocks and operate on the respective token sets. At inference time, samples are drawn by integrating Eq.~\eqref{eq:flow} from $t=0$ to $t=1$ with $u_t$ in place of $v_t$, using an explicit ODE solver. The architecture and inference procedure are visualised in Fig.~\ref{fig:architecture}. Detailed hyperparameter values are reported in Appendix~\ref{app:hparams}.

\begin{figure*}
\resizebox{\textwidth}{!}{%
\begin{tikzpicture}[x=2mm,y=2mm]
  \def\panelx{-6}   
 
  \node[sm,fill=cIn,minimum width=52mm,minimum height=14mm]  (cc) at (10, 9) {$c\in\mathbb{R}^{d_c}$};
  \node[sm,fill=cIn,minimum width=52mm,minimum height=14mm]  (xt) at (10,-9) {$x\in\mathbb{R}^{6}$,\, $t\in[0,1]$};
  \node[opt,fill=cIn,minimum width=52mm,minimum height=14mm] (x0) at (10,-24) {$(x_0^{(i)})_{i=1}^L\in\mathbb{R}^{L \times 6}$};
 
  \node[sm,fill=cProc, minimum width=32mm,minimum height=22mm] (cat) at (36,0)
        {concat\\ $\mathbb{R}^{7+d_c}$};
  \node[sm,fill=cProc, minimum width=32mm,minimum height=22mm] (L0) at (58,0)
        {input\\ layer};
  \node[sm,fill=cModel,minimum width=36mm,minimum height=22mm] (B1) at (80,0)
        {residual\\ block 1};
  \node[sm,draw=cEdge,dashed,minimum width=18mm,minimum height=22mm] (Bd) at (96,0)
        {$\cdots$};
  \node[sm,fill=cModel,minimum width=36mm,minimum height=22mm] (BN) at (114,0)
        {residual\\ block $n$};
  \node[sm,fill=cProc, minimum width=32mm,minimum height=22mm] (L1) at (136,0)
        {output\\ layer};
  \node[sm,fill=cOut,  minimum width=36mm] (U) at (158,0) {$u_t\in\mathbb{R}^{6}$};
 
  \draw[ar] (cc.east) to[out=0,in=155] ([yshift=4mm]cat.west);
  \draw[ar] (xt.east) to[out=0,in=205] ([yshift=-4mm]cat.west);
  \foreach \a/\b in {cat/L0,L0/B1,B1/Bd,Bd/BN,BN/L1,L1/U} \draw[ar] (\a) -- (\b);
 
  \draw[pl] (cc.north) -- (10,20) -- (122,20);
  \foreach \x in {80,96,114} \draw[ar] (\x,20) -- (\x,5.5);
  \node[lbl,anchor=west] at (124,20) {$c$ also gates every block (GLU)};
 
  \node[opt,fill=cModel,minimum width=40mm,minimum height=14mm] (CE) at (36,-24) {coordinate\\ embedding};
  \draw[dar] (x0.east) -- (CE.west);
  \draw[dpl] (CE.east) -- (122,-24);
  \foreach \x in {80,96,114} \draw[dar] (\x,-24) -- (\x,-5.5);
  \node[lbl,anchor=west] at (124,-24)
       {$e\left((x_0^{(i)})_{i=1}^L\right)$ as keys/values\\[4pt] (CA- and Hybrid-CFM only)};
 
  \node[font=\dghead,anchor=west] at (\panelx,26) {(a)};
  \node[font=\dghead,text=black!100,anchor=west] at (\panelx+9, 26)
       {ResNet};
 
  \def\yb{-56}
  \def\dgap{8mm}
  \node[sm,fill=cOut,  minimum width=18mm]  (bi)  at (8,\yb) {in};
  \node[sm,fill=cProc, minimum width=36mm,right=\dgap of bi]  (a1) {activation};
  \node[sm,fill=cProc, minimum width=20mm,right=\dgap of a1]  (fc1) {FC};
  \node[sm,fill=cProc, minimum width=40mm,right=\dgap of fc1] (a2)
        {activation\\ $+$ dropout};
  \node[sm,fill=cProc, minimum width=20mm,right=\dgap of a2]  (fc2) {FC};
  \node[opt,fill=cModel,minimum width=68mm,right=\dgap of fc2] (xa)
        {Cross-attention \\[4pt] $Q=h$; $K,V=e\left((x_0^{(i)})_{i=1}^L\right)$};
  \node[sm,fill=cModel,minimum width=28mm,right=\dgap of xa]  (gl) {GLU};
  \node[blk,circle,inner sep=2pt,font=\dgtext,fill=white,right=\dgap of gl] (sum)
        {$+$};
  \node[sm,fill=cOut,  minimum width=20mm,right=\dgap of sum] (bo) {out};
 
  \foreach \a/\b in {bi/a1,a1/fc1,fc1/a2,a2/fc2,fc2/xa,xa/gl,gl/sum,sum/bo}
    \draw[ar] (\a) -- (\b);
 
  \coordinate (skipY) at ([yshift=10mm]xa.north);
  \draw[ar] (bi.north) -- (bi.north |- skipY) -- (sum.north |- skipY) -- (sum.north);
  \node[lbl,anchor=south] at ([yshift=1mm]xa.north |- skipY)
       {skip connection};
 
  \draw[dar] ([yshift=-9mm]xa.south) -- (xa.south);
  \node[lbl,anchor=north] at ([yshift=-9mm]xa.south) {$e\left((x_0^{(i)})_{i=1}^L\right)$};
  \draw[ar]  ([yshift=-9mm]gl.south) -- (gl.south);
  \node[lbl,anchor=north] at ([yshift=-9mm]gl.south) {$c$};
 
  \begin{scope}[on background layer]
    \node[draw=cGrey,line width=0.6pt,dash pattern=on 4pt off 3pt,
          rounded corners=2pt,fill=black!2,inner sep=0pt,
          fit={(\panelx-2,\yb-20) (\panelx-2,\yb+18) ([xshift=10mm]bo.east)}]
          (det) {};
  \end{scope}
  \node[font=\dghead,anchor=west] at (\panelx,\yb+14) {(b)};
  \node[font=\dghead,text=black!100,anchor=west] at (\panelx+9,\yb+14)
       {Residual block};
  \draw[lead] (B1.south west) -- (40,\yb+18);
  \draw[lead] (B1.south east) -- (120,\yb+18);
 
  \def\yi{-96}
  \foreach \i/\x in {0/20, 1/54, 2/88, 3/122}
    \node[pan] (t\i) at (\x,\yi) {\includegraphics[width=44mm]{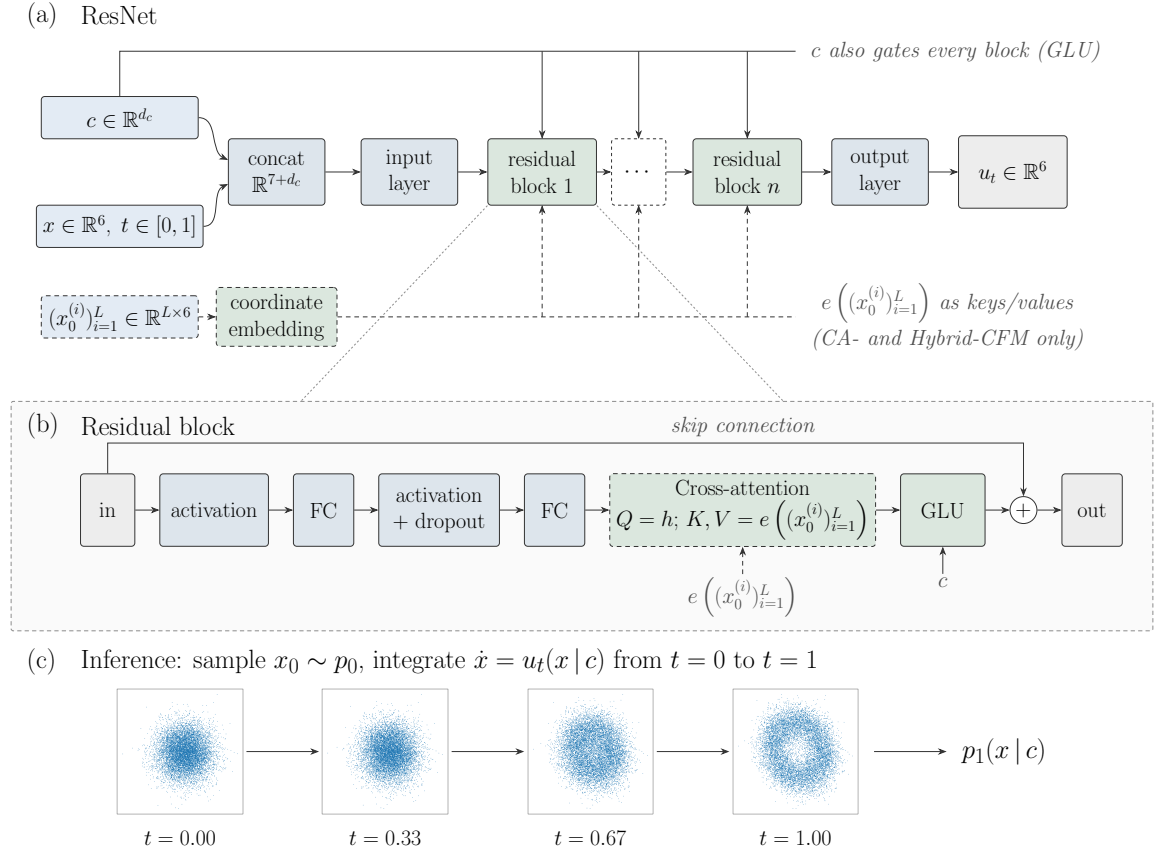}};
  \foreach \a/\b in {t0/t1,t1/t2,t2/t3} \draw[ar] (\a) -- (\b);
  \foreach \i/\v in {0/0.00, 1/0.33, 2/0.67, 3/1.00}
    \node[font=\dgtext,below=3mm of t\i] {$t=\v$};
  \draw[ar] (135,\yi) -- (147,\yi);
  \node[font=\dghead,anchor=west] at (149,\yi) {$p_1(x\,|\,c)$};
  \node[font=\dghead,anchor=west] at (\panelx,\yi+15) {(c)};
  \node[font=\dghead,text=black!100,anchor=west] at (\panelx+9,\yi+15)
       {Inference: sample $x_0\sim p_0$, integrate $\dot{x}=u_t(x\,|\,c)$ from $t=0$ to $t=1$};
 
\end{tikzpicture}}
\caption{(a) General layout of the ResNet: the inputs are the phase-space coordinate \(x \in \mathbb{R}^6\), the machine settings \(c \in \mathbb{R}^{d_c}\) and time \(t \in \left[0, 1\right]\). CA-CFM and Hybrid-CFM additionally take \((x_0^{(i)})_{i=1}^L \in \mathbb{R}^{L \times 6}\), the phase-space coordinates of \(L\) particles, as input. In the case of CA-CFM, these particles are samples from the initial distribution (see Sec.~\ref{sec:ca-cfm}) and for Hybrid-CFM, they are the auxiliary particles (see Sec.~\ref{sec:hybrid}). (b) Layers of each residual block: The input (in) is passed through two interleaved activation and fully-connected (FC) layers. Optionally, a dropout layer~\cite{JMLR:v15:srivastava14a} follows the second activation layer. In the case of CA-CFM and Hybrid-CFM, cross-attention between the internal representation \(h\) and the embedded particle ensemble \(e(x_0^{(i)})_{i=1}^L\) is performed afterwards. Finally, a gated linear unit~\cite{pmlr-v70-dauphin17a} (GLU) injects the machine settings \(c\). A skip connection bypasses these steps, hence the output (out) is the sum of the processed and raw inputs. (c) Inference: sample coordinates \(x_0\) from the initial phase-space distribution \(p_0\) and integrate the vector field \(u_t\) from \(t=0\) to \(t=1\) to obtain samples from the final phase-space distribution \(p_1\). We use the Dormand-Prince 5(4)~\cite{DORMAND198019} integration scheme.}
\label{fig:architecture}
\end{figure*}

\subsection{Evaluation metrics}\label{sec:metric}

Because the model is trained without particle correspondence and outputs an ensemble rather than per-particle predictions, we require a metric that compares two distributions directly. We use the squared maximum mean discrepancy (\(\text{MMD}^2\))~\cite{gretton2012}: let \((x)_{i=1}^N\) be a sample from the ground-truth phase-space distribution and \((y)_{i=1}^M\) a sample generated by the surrogate model, then an estimator of \(\text{MMD}^2\) is given by
\begin{equation}\label{eq:mmd}
\begin{split}
    \mathrm{\widehat{MMD}}^2(x, y) = \frac{1}{N^2} \sum\limits_{i,j = 1}^N k(x_i, x_j) \\ 
    {} - \frac{2}{MN} \sum\limits_{i,j = 1}^{M, N} k(x_i, y_j) + \frac{1}{M^2} \sum\limits_{i,j = 1}^M k(y_i, y_j).
\end{split}
\end{equation}
%
%
We use the radial basis function (RBF) kernel $k(z_1, z_2) = \exp\left(- \gamma \|z_1 - z_2\|^2\right)$ with bandwidth \(\gamma = 1/6\). The \(\text{MMD}\) is a metric on the space of probability distributions with this kernel~\cite{gretton2012}. Henceforth, we simply write \(\text{MMD}^2\) when we refer to the estimator given in Eq.~\ref{eq:mmd}.  As an accelerator-meaningful sanity check, the space-charge experiment of Section~\ref{sec:results:spacecharge} additionally reports the predicted versus true tail population, which probes whether the model captures the halo content. Here, the tail population is defined as the fraction of particles with action \(J > J_{4\sigma}\) where \(J_{4\sigma}\) is the action of a particle at an amplitude of \SI{4}{\sigma}.

\subsection{AI-tools}\label{sec:ai-tools}
Claude Opus 5 and Claude Fable 5 (both Anthropic) were used to generate, debug, clean and annotate code for the companion repository~(~\href{https://gitlab.cern.ch/abt-optics-and-code-repository/simulation-codes/cfmtrack.git}{https://gitlab.cern.ch/abt-optics-and-code-repository/simulation-codes/cfmtrack.git}). Furthermore, these AI-agents were used to prepare, execute and summarise the hyperparameter study (see Appendix~\ref{app:hparams}). Study protocols were drafted and approved by the authors before execution, results were carefully checked and replication pipelines are published in the companion repository.

\section{Results}\label{sec:results}

\subsection{Single-particle tracking in the PS}\label{sec:results:ps10d}

We first apply CFM to single-particle tracking in the PS with a \num{10}-dimensional parameter space, which comprises the normalised strengths of three octupole and three sextupole families together with four AC-dipole chirp parameters (number of turns, amplitude, initial and final frequencies). Training data is generated by sampling \num{1500} points in this space, with \(N = \num{1e4}\)  final-state particles per configuration produced by conventional tracking with Xsuite~\cite{Iadarola:2023fuk}. The trained model is evaluated on 100 distributions generated from previously unseen settings drawn from the same space. The model produces \num{1e4} particles per test configuration, though it can in principle generate any number of particles.

\begin{figure*}
    \centering
    \includegraphics[width=1.0\textwidth]{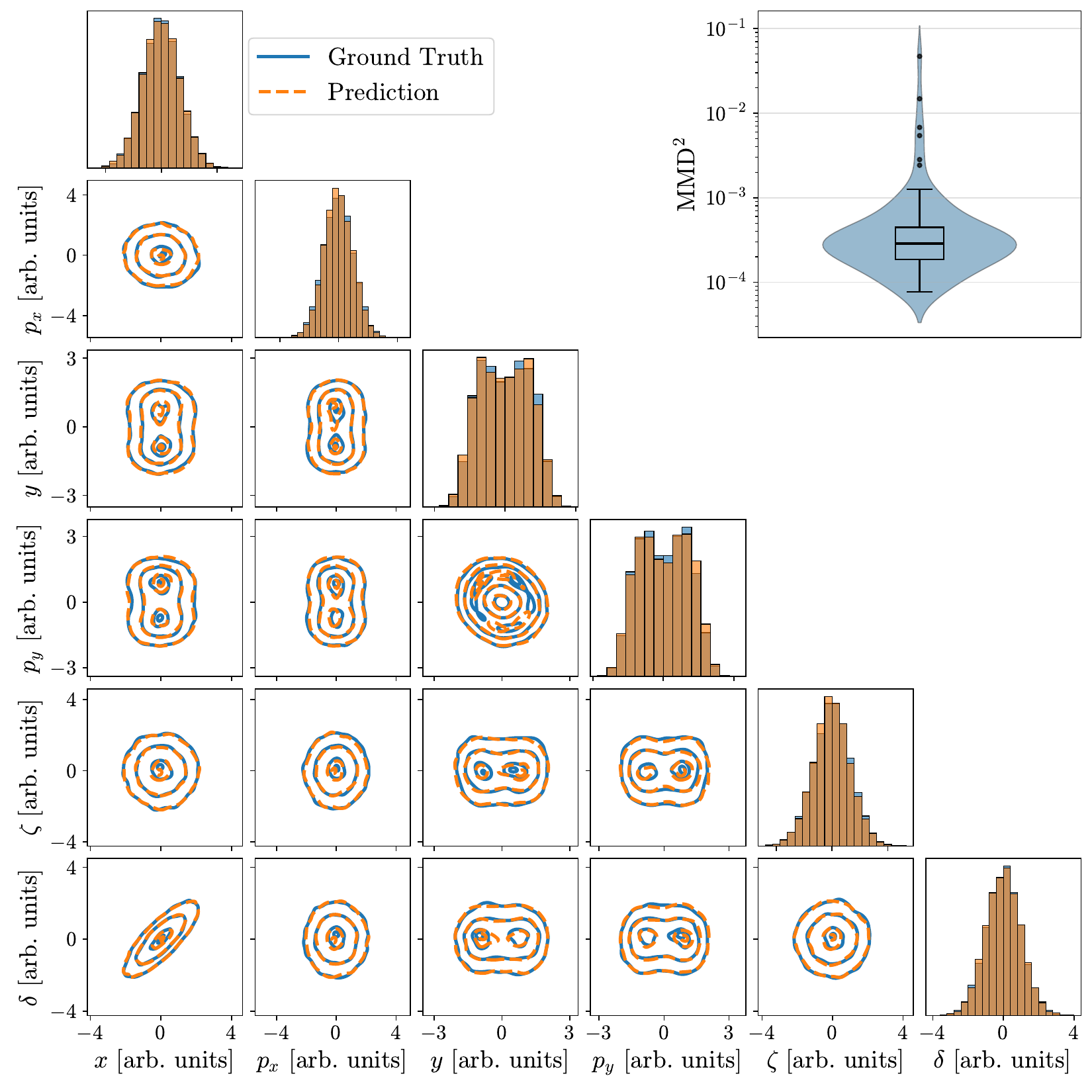}
    \caption{Single representative test configuration from the 10-dimensional PS task (Section~\ref{sec:results:ps10d}). Lower triangle: marginal projections of the six-dimensional phase space, with the CFM prediction (orange) overlaid on the ground truth (blue). Diagonal panels show one-dimensional histograms of each phase-space coordinate. Off-diagonal panels show two-dimensional Gaussian kernel-density-estimation~\cite{parzen1962} contours, with bandwidth selected by Scott's rule~\cite{scott1992}. Top-right inset: histogram of \(\text{MMD}^2\) values over all \num{100} validation distributions. Median \(\text{MMD}^2\) $= \num{3e-4}$, with a long-tailed minority of outlier configurations addressed in Section~\ref{sec:results:hybrid}.}
    \label{fig:cfm_projections}
\end{figure*}

The \(\text{MMD}^2\) distribution over the test set (top-right panel of Fig.~\ref{fig:cfm_projections}) has median $\num{3e-4}$, with the bulk of configurations below $\num{1e-3}$ and a long-tailed minority of outliers, likely caused by the highly non-linear relation between machine settings and final distribution in this example. Section~\ref{sec:results:hybrid} presents a Hybrid-CFM-based mitigation. The lower triangle of Fig.~\ref{fig:cfm_projections} compares the one- and two-dimensional projections of the model prediction (orange) against the ground truth (blue) for a representative test configuration, which agree across all projections. CFM produces a \num{10000}-particle ensemble in approximately \SI{40}{\milli\second}, three orders of magnitude faster than Xsuite (see Table~\ref{tab:inference_time_gpu}).

\subsection{Collective effects: space charge}\label{sec:results:spacecharge}

To test CA-CFM (Section~\ref{sec:ca-cfm}), we simulate distribution-dependent dynamics by performing particle-in-cell space-charge tracking in the PS with Xsuite~\cite{Iadarola:2023fuk}. The initial distributions are mixtures of a matched six-dimensional Gaussian and a halo component, the latter defined by polar coordinates $r_x \sim U(2\sigma_x, 3\sigma_x)$ and $\theta \sim U(0, 2\pi)$ in the normalised horizontal phase-space and $(y, p_y, \zeta, \delta) \sim \mathcal{N}(0, \Sigma)$. The polar coordinates are converted into canonical coordinates $(x, p_x)$, which are independent of $(y, p_y, \zeta, \delta)$. Evenly spaced mixture weights between 0 and 1 yield a one-parameter family of initial distributions. A mixture weight of \(0\) yields purely the halo component. Conversely, the matched Gaussian beam is obtained with a mixture weight of \(1\). Each distribution is tracked for 256 turns with 50 space-charge kicks per turn. Of the 98 distributions, 60 are used for training and 38 for testing.

\begin{figure*}
    \centering
    \includegraphics[width=1.0\textwidth]{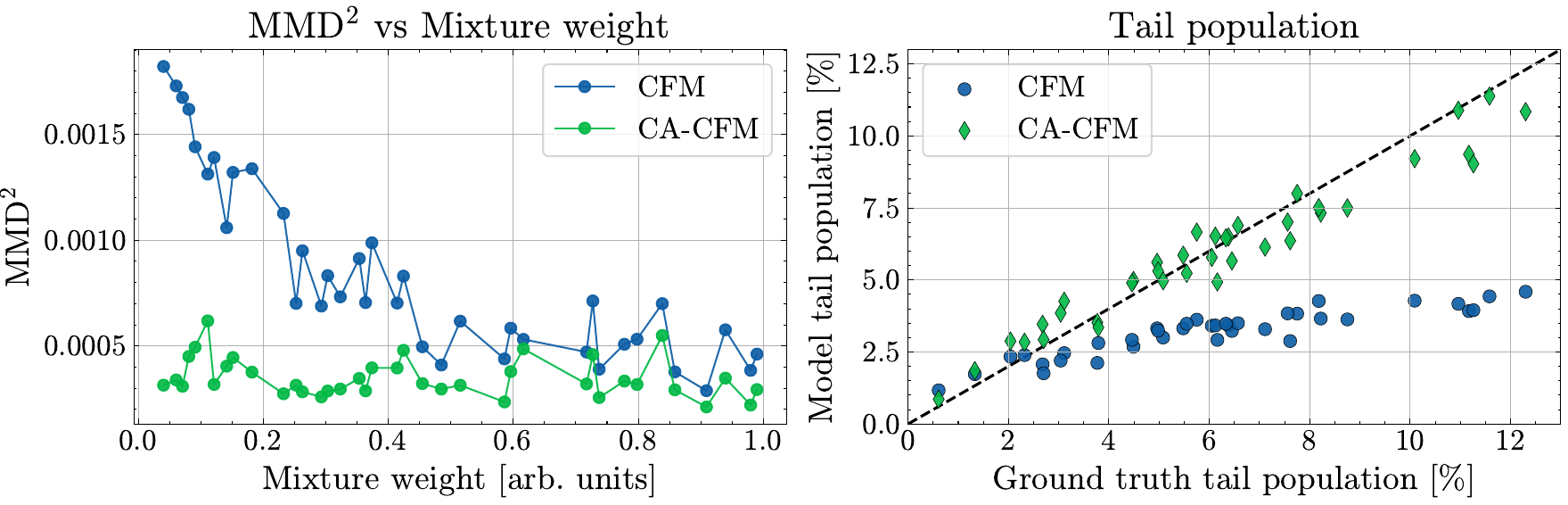}
    \caption{Left: \(\text{MMD}^2\) between the predicted and ground-truth final phase-space distributions for the PS space-charge experiment (Section~\ref{sec:results:spacecharge}), as a function of the Gaussian--halo mixture weight. Vanilla CFM (blue) degrades at small mixture weights, while CA-CFM (green) maintains a consistent \(\text{MMD}^2\) across the full mixture range. Right: model prediction versus ground truth for the fraction of particles in the horizontal tail. The tail population is large due to strongly non-linear dynamics and space-charge effects. Vanilla CFM (blue) systematically underestimates the tails, while CA-CFM (green) agrees well with the ground truth across the full range.}
    \label{fig:space_charge_metrics}
\end{figure*}

For both models, the true initial particle coordinates serve as the source distribution $x_0$ at training and as starting points at inference. For CA-CFM, they additionally serve as cross-attention tokens. Performance as a function of mixture weight is shown in the left panel of Fig.~\ref{fig:space_charge_metrics}: vanilla CFM degrades at small mixture weights, while CA-CFM maintains a consistent \(\text{MMD}^2\) across the full range. Small mixture weights correspond to a large halo content in the original distribution, where the stronger non-linear dynamics may explain the degraded performance of vanilla CFM. The right panel of Fig.~\ref{fig:space_charge_metrics} compares the predicted against the true tail population of the horizontal plane. Vanilla CFM fails to accurately predict strongly populated tails. On the other hand, CA-CFM tracks the true tail population well across the entire range. CA-CFM evolves a \num{10000}-particle ensemble in approximately~\SI{0.4}{\second}, three orders of magnitude faster than Xsuite (see Table~\ref{tab:inference_time_gpu}).

\subsection{Data efficiency with Hybrid-CFM}\label{sec:results:hybrid}

We evaluate Hybrid-CFM (Section~\ref{sec:hybrid}) on the same 10-dimensional single-particle PS task as in Section~\ref{sec:results:ps10d}, varying both the training-set size and the number of auxiliary particles. As a baseline, we train a vanilla CFM model. Then, we train Hybrid-CFM models with $N_\text{aux} \in \{10, 50, 100, 200, 500\}$. Each setting is repeated for training-set sizes of 200, 500 and 1500 distributions ($N = 10\,000$ particles per distribution). The number of training distributions corresponds directly to the upfront tracking-simulation cost, since one conventional tracking run is required per distribution.

\begin{figure*}
    \centering
    \includegraphics[width=1.0\textwidth]{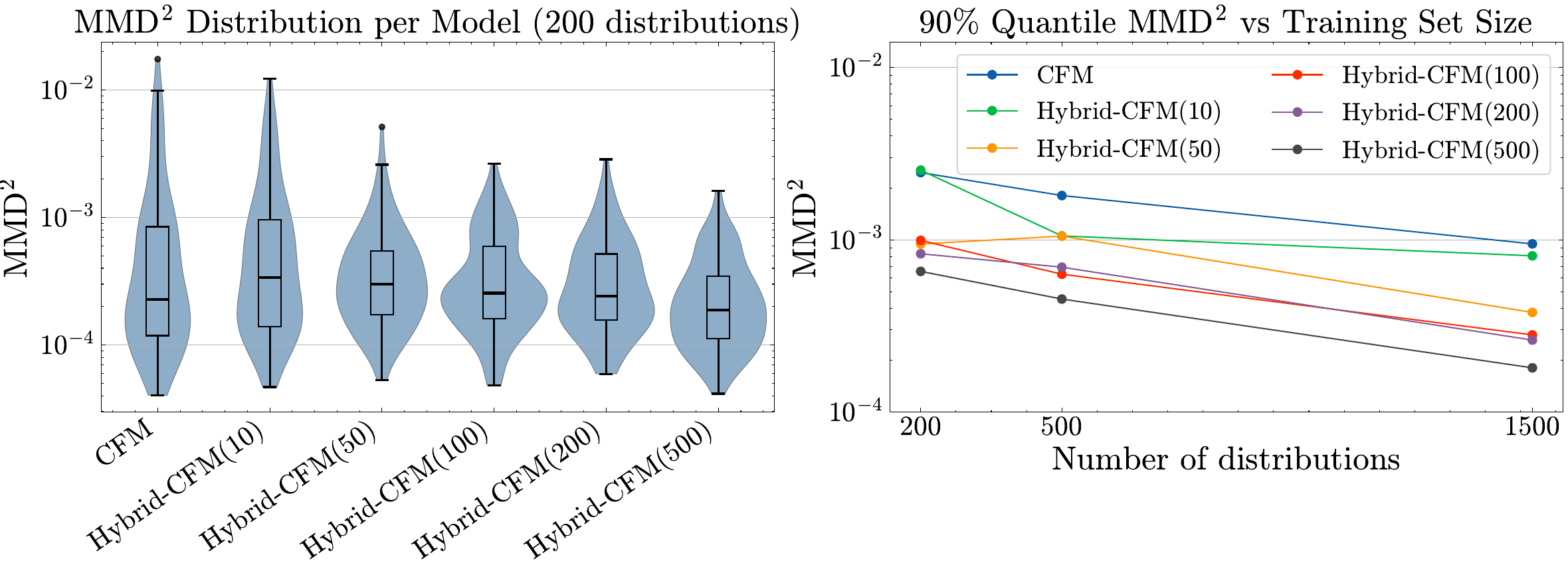}
    \caption{CFM and Hybrid-CFM models trained with $N_\text{aux} \in \{10, 50, 100, 200, 500\}$ auxiliary particles on training sets of 200, 500 and 1500 distributions ($N = 10\,000$ particles per distribution). Left: per-test-configuration \(\text{MMD}^2\) for the 200-distribution training set; Hybrid-CFM compresses the outlier tail for $N_\text{aux} \gtrsim 50$, with little effect at $N_\text{aux} = 10$. Right: 90th-percentile \(\text{MMD}^2\) versus training-set size; Hybrid-CFM matches the CFM's 90th-percentile \(\text{MMD}^2\) with substantially fewer training distributions and improves the 90th-percentile \(\text{MMD}^2\) by up to a factor of four on equally-sized training sets.}
    \label{fig:guidance_metrics}
\end{figure*}

Figure~\ref{fig:guidance_metrics} summarises the results. The left panel shows \(\text{MMD}^2\) for the different models trained on the 200-distribution training set: the vanilla CFM performs well in the median but exhibits a long tail of outlier configurations. With $N_\text{aux} = 10$, Hybrid-CFM provides little benefit, presumably because the auxiliary signal is too noisy to anchor the predicted distribution. For $N_\text{aux} \gtrsim 50$, however, Hybrid-CFM substantially compresses the outlier tail, with diminishing returns above $N_\text{aux} \approx 100$. The right panel shows the 90th-percentile \(\text{MMD}^2\) as a function of training-set size for each $N_\text{aux}$: Hybrid-CFM improves the worst-case \(\text{MMD}^2\) by up to a factor of four. In particular, Hybrid-CFM with $N_\text{aux} = 100$ trained on 200 distributions matches the 90th-percentile \(\text{MMD}^2\) of the vanilla CFM trained on 1500 distributions -- an up to $7.5\times$ reduction in upfront tracking cost at matched worst-case (90th-percentile) accuracy. The inference time of this method for $N = 10\,000$ particles is dominated by the tracking cost of the auxiliary particles and the total speed-up on GPU is negligible (see Table~\ref{tab:inference_time_gpu}). Since this auxiliary cost is independent of \(N\), it is amortized as \(N\) grows, and we therefore expect increasing gains over conventional tracking. On CPU, a $47\times$ speed-up was achieved with \(N_\text{aux}=50\).  

\section{Discussion}\label{sec:discussion}
The surrogate models proposed here treat particle tracking as a distributional map, collapsing the per-turn motion of $N$ particles into one learned transport. This sidesteps the dominant cost of long-term tracking in circular machines, at the price of giving up particle-level resolvability and any explicit physics prior. The remainder of this section discusses the resulting compute trade-offs, the limitations this framing imposes, and future research directions.

The total compute cost of the approach has three components: the upfront cost of generating the training set by conventional tracking, the cost of training the CFM model, and the cost of per-query inference (and, for Hybrid-CFM, of the per-query auxiliary tracking). Pure conventional tracking pays a large cost at every query, whereas the CFM and CA-CFM models amortise this cost upfront and offer near-real-time inference. Whether this trade-off pays off therefore depends strongly on the cost of a single conventional query. For single-pass systems such as linacs or transfer lines, where a particle traverses each element once, direct tracking is already cheap and the upfront investment is unlikely to be recovered. The picture changes for circular machines: tracking many turns and large ensembles of particles makes each query expensive enough that the upfront cost is repaid after comparatively few forward evaluations. Hybrid-CFM trades a smaller upfront tracking cost for a per-query auxiliary-tracking overhead, which is favourable when training data is scarce and the auxiliary cost remains much smaller than tracking the full $N$ particles.

The surrogate is not symplectic by construction. Enforcing symplecticity would require the learned vector field itself to be symplectic, which in the limit reduces to a Hamiltonian neural network operating at the per-particle level -- in essence, the class of physics-informed and integrator-based surrogates discussed in Section~\ref{sec:intro}. Such methods must explicitly resolve the non-linear, possibly time-dependent motion of every particle over every turn, which becomes the dominant cost for long-term tracking in circular machines~\cite{remta2025}. By collapsing that motion into a single distributional map, the present approach sidesteps this bottleneck, but is therefore unsuited to studies that require particle-level accuracy. Another consequence of omitting explicit physics priors is degraded accuracy in regions of the parameter space where training data is sparse. This is visible in the heavy-tailed \(\text{MMD}^2\) distribution on the 10-dimensional PS task in Section~\ref{sec:results:ps10d}. Two mitigations are available. The brute-force option is to enlarge the training set, which is expensive because each data-point requires a full tracking simulation. Hybrid-CFM is the more economical alternative: tracking $N_\text{aux} \ll N$ auxiliary particles per configuration is cheap, and the resulting tokens anchor the prediction in the physics of the specific configuration. As shown in Section~\ref{sec:results:hybrid}, this yields up to a $7.5\times$ reduction in upfront tracking cost at matched worst-case accuracy.

Three further directions deserve mention. First, uncertainty quantification: a downstream optimiser using the surrogate needs to know where the prediction is trustworthy. Model ensembles~\cite{3295222.3295387, 3454287.3455466} would address this issue but incur additional training and inference cost. Another promising approach is conformal prediction~\cite{10.1561/2200000101, kladny2025}. Second, incorporation of measured data: instrumentation on real accelerators delivers distribution-level rather than per-particle observables, often as lower-dimensional projections of the full six-dimensional phase space (e.g.\ two-dimensional screen images, transverse profiles, tomographic reconstructions). The distributional nature of CFM lends itself naturally to such data sources, opening a path to surrogates that combine simulation-based and measurement-based training. Third, system identification~\cite{Tejero-Cantero2020, boelts2025sbi, deistler2025simulationbasedinferencepracticalguide, NEURIPS2023_3663ae53}: the true machine parameters might be inferred by matching the surrogate output against measured beam distributions. Inverse problems of this kind are inherently iterative — a Bayesian or gradient-based sampler would need to evaluate the forward model many thousands of times while exploring the parameter space — which makes them impractical when each evaluation requires a full tracking simulation. A surrogate that reproduces the tracked distribution in milliseconds could bring such studies within reach, and its differentiability would in principle allow gradients with respect to the machine parameters to be propagated directly to the optimiser.

\section{Conclusion}\label{sec:conclusion}

We presented a generative surrogate-modelling approach for particle tracking based on CFM. The model is trained on tracking simulations parametrised by machine settings Once trained, it produces final phase-space distributions up to three orders of magnitudes faster than the conventional tracking code. We introduced two extensions: CA-CFM, which captures distribution-dependent dynamics such as collective effects, and Hybrid-CFM, which uses a small ensemble of conventionally tracked auxiliary particles to mitigate the curse of dimensionality in high-dimensional parameter spaces. On a 10-dimensional task in the PS, CFM reproduces final-state phase-space distributions with a median \(\text{MMD}^2\) of $\num{3e-4}$. Hybrid-CFM achieves matched worst-case accuracy with up to a $7.5\times$ reduction in upfront tracking cost. CA-CFM captures space-charge-induced halo dynamics where vanilla CFM degrades. 

Because the surrogate models are differentiable with respect to the machine settings, they can be plugged directly into gradient-based optimisation loops in the parameter space, opening efficient design and online-control workflows that are impractical with conventional tracking. With inference times between 0.04 and 0.4 seconds, the models are fast enough to serve as a building blocks for virtual accelerators and digital twins. The flexibility of the underlying framework -- arbitrary conditioning, distribution-level training, and compatibility with conventional tracking methods -- suggests broad applicability across accelerator design, optimisation, and online-control tasks.

Beyond particle accelerators, the approach naturally extends to any setting where ensembles of particles evolve under (stochastic) differential equations and the distribution itself is the quantity of interest, such as molecular dynamics and plasma physics.

\begin{acknowledgments}
M.R. designed and carried out the studies, developed the software, and wrote the manuscript. Y.D. and F.V. supervised the work and contributed ideas and feedback in discussions. F.V. contributed in the revision of the manuscript. We acknowledge the substantive usage of AI-tools during this study, as disclosed in Sec.~\ref{sec:ai-tools}.
\end{acknowledgments}

\section*{Data and code availability}
Trained model checkpoints~\cite{models2026} and the simulation datasets~\cite{datasets2026} used in this work are publicly available on Zenodo. Training and inference code is publicly available on GitLab (\hyperlink{https://gitlab.cern.ch/abt-optics-and-code-repository/simulation-codes/cfmtrack.git}{https://gitlab.cern.ch/abt-optics-and-code-repository/simulation-codes/cfmtrack.git}).

\appendix

\section{Hyperparameter studies}\label{app:hparams}
This section presents the impact of certain hyperparameters on model performance. All hyperparameter studies were performed on the \num{200}-distribution training set from the PS single-particle tracking problem (see Sec.~\ref{sec:results:ps10d}). Performance is measured on the validation set (100 distributions), except for the final evaluation, which is performed on the test set (332 distributions). 
 
\begin{figure*}
    \centering
    \includegraphics[width=1.0\textwidth]{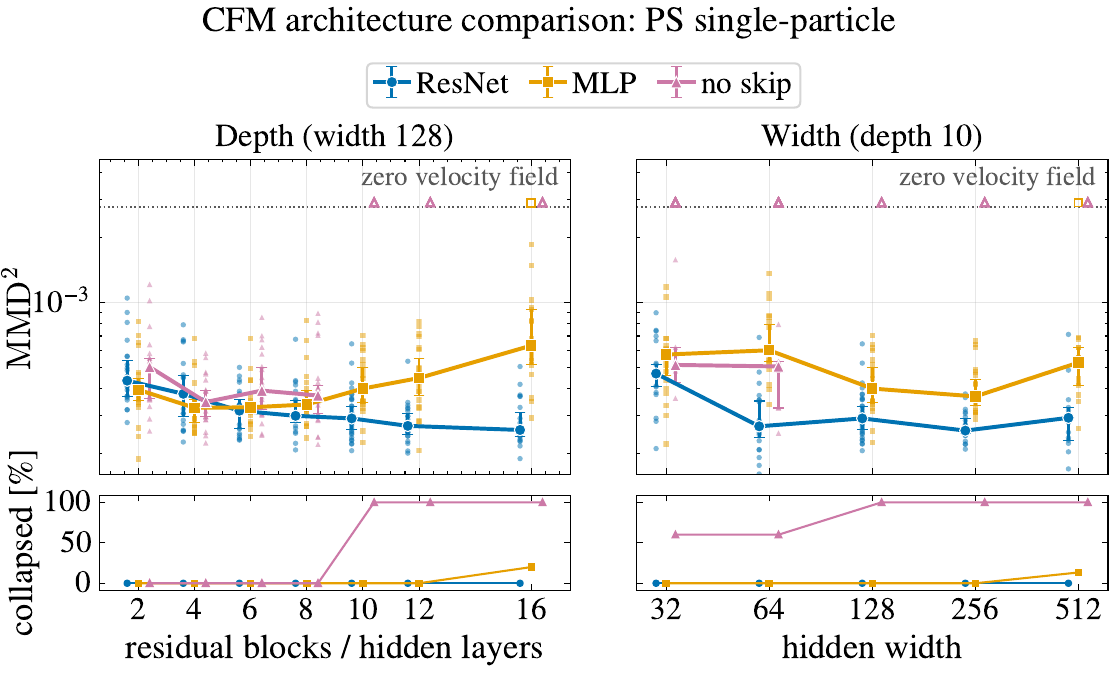}
    \caption{Three architectures benchmarked on the PS single-particle tracking problem: ResNet (blue), MLP (orange), and the same ResNet without skip connections (magenta). All models are trained on the \num{200}-distribution training set. Each configuration is trained with up to \num{20} independent seeds, and each seed is scored by the median \(\text{MMD}^2\) over \num{20} configurations from the validation set. Top: markers show the median score over non-collapsed seeds, with \SI{95}{\percent} bootstrap confidence intervals on that median. Individual seeds are shown as faint points. Bottom: fraction of seeds that collapsed during training, defined as \(\text{MMD}^2\) exceeding the smallest value of 100 evaluations of a model with identically zero velocity field, whose output is simply its initial noise ensemble. Left: width fixed at \num{128} nodes, depth varied from \numrange{2}{16}. Right: depth fixed at \num{10}, width varied from \numrange{32}{512}. Markers belonging to the same group are slightly offset horizontally for visual clarity.}
    \label{fig:hpo_architecture}
\end{figure*}

\emph{Network architecture}. We obtained our results (see Sec.~\ref{sec:results}) using residual networks~\cite{he2016}, which each consist of 10 residual blocks, as backbone of our models. The layout is shown in Fig.~\ref{fig:architecture} and follows the implementation in Ref.~\cite{Stimper2023}, where each residual block consists of two non-linear activations (rectified linear unit~\cite{hinton2010} (ReLU)) interleaved with two linear layers of 128 neurons. Each block is wrapped with a skip connection. Conditioning inputs are injected with a gated linear unit~\cite{pmlr-v70-dauphin17a} in each block. For our CA-CFM and our Hybrid-CFM, we add a single-head attention block~\cite{vaswani2017, pmlr-v97-lee19d} to each residual block. The number of trainable weights is stated in Table~\ref{tab:inference_time_gpu} for each model. \\
We compare this network architecture with two alternatives, sharing all remaining hyper-parameters with the published configuration (see Tables~\ref{tab:hparams}~and~\ref{tab:batch_size}): the same network without skip connections, and a plain multi-layer perceptron (MLP) receiving the context once at the input. The three architectures were compared on a grid over depth (2–16 residual blocks / hidden layers at width 128) and width (32–512 at depth 10). Each of the 33 configurations was trained from 20 independent seeds (weight initialisation, data ordering, and path noise all re-drawn) — reduced to 15 for the three width-512 cells and to 5 for the five no-skip cells in which all seeds collapse — amounting to 585 runs in total. The \(\text{MMD}^2\)-performance is shown in Fig.~\ref{fig:hpo_architecture}. The three architectures perform similarly when the network is shallow (6 or less hidden layers). As the depth is increased, ResNet's performance improves slightly, whilst the MLP degrades and the ResNet without skip connections even collapses. The latter is likely caused by the GLU modulation. Hence, skip connections facilitate deeper architectures, which is consistent with previous studies~\cite{he2016}. On the other hand, the performance curve for different widths is flat, only 32 neurons perform significantly worse across architectures. A ResNet with depth 10 and width 64 provides the same performance as our published baseline with a factor \(3.8\) less weights. 

\emph{Activation function}. Our baseline architecture uses ReLU as activation function. We compare the performance against Tangens hyperbolicus (Tanh), gaussian error linear unit~\cite{hendrycks2023gaussianerrorlinearunits} (GELU) and sigmoid linear unit~\cite{hendrycks2023gaussianerrorlinearunits, elfwing2017sigmoidweightedlinearunitsneural, ramachandran2017searchingactivationfunctions} (SiLU). Only Tanh differs significantly from ReLU (\(1.34\times\) worse, \SI{95}{\percent} confidence interval excludes unity), GELU and SiLU are interchangeable with the baseline. The results are summarised in Table~\ref{tab:activations}.

\begin{table}
    \centering
    \caption{Activation function comparison at the published architecture
    ($10\times128$ ResNet, all other hyper-parameters at their published
    values). Each activation is trained with $n$ independent seeds. Each seed
    is scored by the median \(\text{MMD}^2\) over 20 validation distributions.
    \emph{Median} is the median score over seeds in units of $10^{-4}$.
    \emph{Ratio} is the ratio of medians relative to the ReLU reference, with
    the \SI{95}{\percent} bootstrap confidence interval over training seeds in brackets.}
    \label{tab:activations}
    \begin{tabular*}{\columnwidth}{@{\extracolsep{\fill}}l S[table-format=2.0] S[table-format=1.2] S[table-format=1.2] l @{}}
        \toprule
        Activation & {$n$} & {Median [$10^{-4}$]} & {Ratio} & {95\,\% CI} \\
        \midrule
        ReLU (ref.) & 30 & 2.91 & 1.00 & {--}          \\
        GELU        & 20 & 2.80 & 0.96 & [0.84, 1.12]  \\
        SiLU        & 20 & 3.11 & 1.07 & [0.91, 1.31]  \\
        Tanh        & 20 & 3.89 & 1.34 & [1.08, 1.57]  \\
        \bottomrule
    \end{tabular*}
\end{table}

\begin{figure}
    \centering
    \includegraphics[width=1.0\columnwidth]{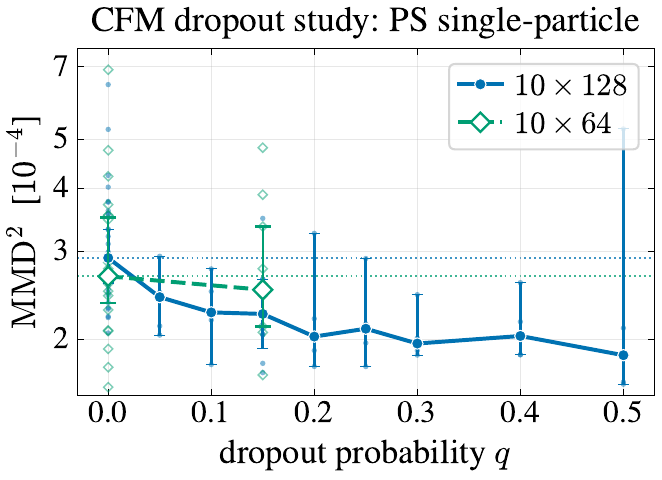}
    \caption{Performance comparison of different dropout probabilities \(q\) for the published architecture (blue) and the smaller variant (green). Markers show the median \(\text{MMD}^2\) over the seeds, with \SI{95}{\percent} bootstrap confidence intervals on that median. Individual seeds are shown as faint points.}
    \label{fig:hpo_dropout}
\end{figure}

\emph{Dropout.} Each residual block incorporates a dropout layer (see Fig.~\ref{fig:architecture}), which randomly sets input features to zero with a probability \(q\)~\cite{JMLR:v15:srivastava14a}. Our baseline architecture does not use dropout, i.e. \(q=0\). Increasing \(q\) for the full-size model (depth 10, width 128) improves the validation \(\text{MMD}^2\) monotonically up to \(q \approx 0.15\), after which the curve flattens (see Fig.~\ref{fig:hpo_dropout}). At \(q = 0.15\) the gain is \num{0.77} (bootstrapped \SI{95}{\percent} confidence interval: \(\left[0.63, 0.96\right]\)) relative to the unregularised model on validation, and \num{0.84} \(\left[0.80, 0.87\right]\) on the held-out test set. The gain does not transfer to the smaller model (depth 10, width 64): there, the same dropout leaves the score almost unchanged (gain \num{0.94} \(\left[0.69, 1.31\right]\)), consistent with a regularisation effect that the reduced capacity no longer requires. Dropout and the reduction in model size are therefore alternatives rather than a combination. Dropout is applied during training only, sampling always uses the deterministic network.

\begin{figure*}
    \centering
    \includegraphics[width=1.0\textwidth]{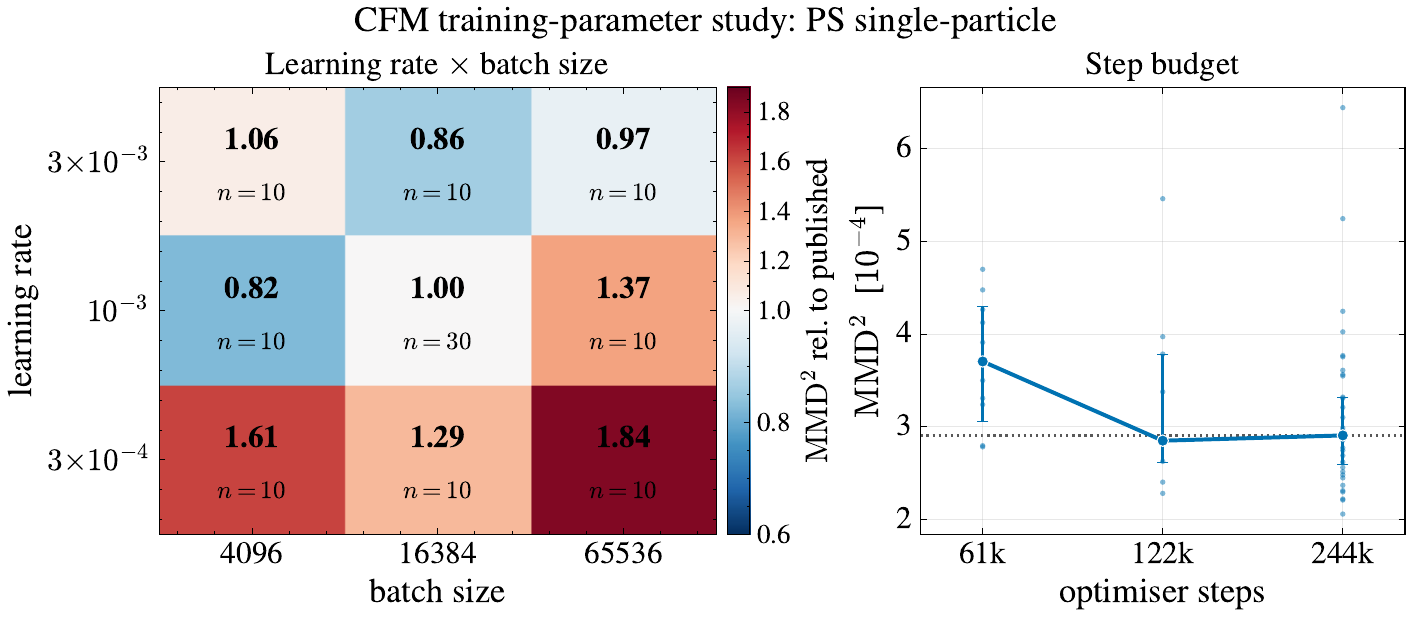}
    \caption{Left: Median of \(\text{MMD}^2\) over multiple seeds for different combinations of learning rate and batch size, relative to the baseline hyperparameters (learning rate = \num{1e3} and batch size = 16384). Right: \(\text{MMD}^2\) versus number of optimiser steps for the baseline architecture. Markers show the median \(\text{MMD}^2\) over the seeds, with \SI{95}{\percent} bootstrap confidence intervals on that median. Individual seeds are shown as faint points.}
    \label{fig:hpo_training}
\end{figure*}

\emph{Training}. We use AdamW with cosine-annealing as the optimiser. All parameters not listed in Table~\ref{tab:hparams} are kept at their PyTorch defaults. Batch size and epochs for the published checkpoints are given in Table~\ref{tab:batch_size}. The batch size refers to the number of particles to ensure comparability over all models. For all models, expect CFM on the single-particle task, the particles are grouped per machine configuration into sets of \num{2000} particles. The number of training epochs depends on the training dataset size, as smaller datasets require more passes to converge. We did not perform early-stopping. \\
We studied the impact of different learning rate and batch size combinations on the \(\text{MMD}^2\)-performance, using the baseline architecture (depth \num{10}, width \num{128}). The results are shown in the left panel of Fig.~\ref{fig:hpo_training}: reducing the learning rate to \num{3e-4} from \num{1e-3} negatively impacts model performance across all batch sizes. Larger batch sizes also require larger learning rates to perform well. Furthermore, we evaluated the influence of overall training budget, measured in optimiser steps. We found, that the optimiser steps can be halved (\num{122e3} steps instead of \num{244e3}) without compromising performance, but reducing to a quarter is harmful (see right panel of Fig.~\ref{fig:hpo_training}).

\begin{table}[h]
\centering
\caption{Training hyperparameters for the published checkpoints, identical for all three tasks presented in Sec.~\ref{sec:results}.}
\label{tab:hparams}
\begin{tabular*}{\columnwidth}{@{\extracolsep{\fill}}ll}
\toprule
Parameter & Value \\
\midrule
Optimiser & AdamW \\
LR schedule & cosine annealing \\
Initial learning rate & \num{1e-3} \\
Final learning rate & \num{1e-4} \\
\bottomrule
\end{tabular*}
\end{table}

\begin{table*}
    \centering
    \caption{Batch size and number of epochs per task and model. Batch size is given in number of particles, which are grouped per machine configuration into sets of 2000 particles. CFM on the single-particle task is the exception: particles are not grouped into sets. Where multiple values are listed, they correspond in order.}
    \label{tab:batch_size}
    \begin{tabular*}{\textwidth}{@{\extracolsep{\fill}}lcccc}
        \toprule
        Task & Method & Training set size & Batch size & Epochs\\
        \midrule
        \multirow{2}{*}{\shortstack[l]{Single-particle\\ dynamics}}
            & CFM                        & 200, 500, 1500 & 16384 & 2000, 2000, 500\\
            & Hybrid-CFM                 & 200, 500, 1500 & 64000 (32) & 5000, 2000, 2000\\
        \midrule
        \multirow{2}{*}{Space-charge}
            & CFM                        & 60 & 64000 (32) & 2000\\
            & CA-CFM                     & 60 & 64000 (32) & 2000\\
        \bottomrule
    \end{tabular*}
\end{table*}

\emph{Final evaluation.} Finally, the most promising hyperparameter-combinations are evaluated on the test set for 10 seeds: depth \num{10}, width \num{128}, \(q = 0.00\) (the published baseline); depth \num{10}, width \num{128}, \(q = 0.15\) and depth \num{10}, width \num{64}, \(q = 0.00\). These models are also compared to the published checkpoint, which is a single training run with the baseline configuration. The results are presented in Table~\ref{tab:final_eval}. The baseline and the \(3.8
\times\) smaller model perform equally, allowing faster training and inference. The best performance is achieved by the large model with dropout regularisation.

\begin{table}
    \centering
    \caption{Final evaluation on the held-out test set (332 distributions),
    performed once after all hyper-parameter decisions were frozen. Each
    configuration is retrained with $n$ independent seeds. Each seed is scored
    by the median $\text{MMD}^2$ over all test distributions. \emph{Median} is
    the median score over seeds in units of $10^{-4}$, \emph{Params} the
    number of trainable weights in Millions.The published model is a single training run.}
    \label{tab:final_eval}
    \begin{tabular*}{\columnwidth}{@{\extracolsep{\fill}}l S[table-format=2.0] S[table-format=1.3] S[table-format=1.2] @{}}
        \toprule
        Configuration & {$n$} & {Params [M]} & {Median [$10^{-4}$]} \\
        \midrule
        $10\times128$, $q=0.00$   & 10 & 0.347 & 2.28 \\
        $10\times128$, $q=0.15$   & 10 & 0.347 & 1.92  \\
        $10\times64$,\,\,\, $q=0.00$    & 10 & 0.092 & 2.38 \\
        \midrule
        Published model       &  1 & 0.347 & 1.94  \\
        \bottomrule
    \end{tabular*}
\end{table}

\section{Inference}\label{app:inference}
For inference, we use Dormand-Prince 5(4)~\cite{DORMAND198019}, implemented in the torchdyn~\cite{politorchdyn} and the torchdiffeq~\cite{torchdiffeq} package, to integrate the neural vector fields. We use absolute and relative tolerances of \num{1e-4}. For the Gaussian probability paths, we set \(\sigma = 0.01\). Table~\ref{tab:inference_time_gpu} states the inference times on the two tasks for the different methods on GPU and Table~\ref{tab:inference_time_cpu} reports the respective values on CPU. Timings were obtained on cluster nodes, using a single NVIDIA H100 GPU (GPU measurements), and with 4 allocated CPU cores (CPU measurements).

\begin{table*}
    \centering
    \caption{Inference time per distribution of $N = 10\,000$ particles on GPU, averaged over the validation set. The number of auxiliary particles for Hybrid-CFM is stated in brackets in \emph{Method}. \emph{Params} states the number of trainable weights in units of Millions. \emph{Model} is the duration of the network forward pass in seconds. \emph{Aux} is the duration of the auxiliary-particle tracking with Xsuite required by Hybrid-CFM. \emph{Total} is their sum. \emph{Speed-up} is relative to conventional tracking with Xsuite.}
    \label{tab:inference_time_gpu}
    \begin{tabular*}{\textwidth}{@{\extracolsep{\fill}}ll S[table-format=3.3] S[table-format=3.3] S[table-format=1.2e1] S[table-format=1.2e1] S[table-format=1.2e1] @{}}
        \toprule
        Task & Method & {Params [M]} & {Model [s]} & {Aux [s]} & {Total [s]} & {Speed-up} \\
        \midrule
        \multirow{7}{*}{\shortstack[l]{Single-particle\\ dynamics}}
            & Xsuite                     & {--}  & 150.632 & {--}   & 1.51e2   & {--}   \\
            & CFM                        & 0.347 & 0.041   & {--}   & 4.09e-2  & 3.69e3  \\
            & Hybrid-CFM(10)             & 1.008 & 0.106   & 1.26e2 & 1.26e2   & 1.20e0  \\
            & Hybrid-CFM(50)             & 1.008 & 0.109   & 1.40e2 & 1.40e2   & 1.08e0  \\
            & Hybrid-CFM(100)            & 1.008 & 0.112   & 1.31e2 & 1.31e2   & 1.15e0  \\
            & Hybrid-CFM(200)            & 1.008 & 0.099   & 1.35e2 & 1.35e2   & 1.12e0  \\
            & Hybrid-CFM(500)            & 1.008 & 0.196   & 1.41e2 & 1.41e2   & 1.07e0  \\
        \midrule
        \multirow{3}{*}{Space-charge}
            & Xsuite                     & {--}  & 509.429 & {--}   & 5.09e2   & {--}   \\
            & CFM                        & 0.333 & 0.042   & {--}   & 4.22e-2 & 1.2e4  \\
            & CA-CFM                     & 0.995 & 0.404   & {--}   & 4.04e-1 & 1.3e3  \\
        \bottomrule
    \end{tabular*}
\end{table*}


\begin{table*}
    \centering
    \caption{Inference time per distribution of $N = 10\,000$ particles on CPU, averaged over the validation set. The number of auxiliary particles for Hybrid-CFM is stated in brackets in \emph{Method}. \emph{Params} states the number of trainable weights in units of Millions. \emph{Model} is the duration of the network forward pass in seconds. \emph{Aux} is the duration of the auxiliary-particle tracking with Xsuite required by Hybrid-CFM. \emph{Total} is their sum. \emph{Speed-up} is relative to conventional tracking with Xsuite.}
    \label{tab:inference_time_cpu}
    \begin{tabular*}{\textwidth}{@{\extracolsep{\fill}}ll S[table-format=1.3] S[table-format=1.2e1] S[table-format=1.2e1] S[table-format=1.2e1] S[table-format=1.1e1] @{}}
        \toprule
        Task & Method & {Params [M]} & {Model [s]} & {Aux [s]} & {Total [s]} & {Speed-up} \\
        \midrule
        \multirow{7}{*}{\shortstack[l]{Single-particle\\ dynamics}}
            & Xsuite                     & {--}  & 1.19e4  & {--}   & 1.19e4   & {--}   \\
            & CFM                        & 0.347 & 6.70e0  & {--}   & 6.70e0   & 1.8e3  \\
            & Hybrid-CFM(10)             & 1.008 & 6.86e0  & 6.35e1 & 7.04e1   & 1.7e2  \\
            & Hybrid-CFM(50)             & 1.008 & 7.20e0  & 2.46e2 & 2.53e2   & 4.7e1  \\
            & Hybrid-CFM(100)            & 1.008 & 7.52e0  & 4.72e2 & 4.80e2   & 2.5e1  \\
            & Hybrid-CFM(200)            & 1.008 & 1.63e1  & 9.19e2 & 9.35e2   & 1.3e1  \\
            & Hybrid-CFM(500)            & 1.008 & 4.76e1  & 2.19e3 & 2.24e3   & 5.3e0  \\
        \midrule
        \multirow{3}{*}{Space-charge}
            & Xsuite                     & {--}  & 3.44e3  & {--}   & 3.44e3   & {--}   \\
            & CFM                        & 0.333 & 6.59e0  & {--}   & 6.59e0   & 5.2e2  \\
            & CA-CFM                     & 0.995 & 9.74e1  & {--}   & 9.74e1   & 3.5e1  \\
        \bottomrule
    \end{tabular*}
\end{table*}

\FloatBarrier
\bibliographystyle{abbrvnat}
\bibliography{References}

\end{document}